\documentclass[aps,prb,superscriptaddress,twocolumn]{revtex4-1}
\usepackage{graphicx}
\usepackage{natbib}

\begin{document}

\title{Magnetic Order and Fluctuations in the Presence of Quenched Disorder in the Kagome Staircase System (Co$_{1-x}$Mg$_{x}$)$_{3}$V$_{2}$O$_{8}$}
\author{K. Fritsch}
\affiliation{Department of Physics and
Astronomy, McMaster University, Hamilton, Ontario, L8S 4M1, Canada}

\author{Z. Yamani}
\affiliation{Canadian Neutron Beam Centre, National
Research Council, Chalk River Laboratories, Chalk River, Ontario, K0J
1P0, Canada}

\author{S. Chang}
\affiliation{NIST Center for Neutron Research, NIST, Gaithersburg, Maryland 20899-8102, USA}

\author{Y. Qiu}
\affiliation{NIST Center for Neutron Research, NIST, Gaithersburg, Maryland 20899-8102, USA}
\affiliation{Department of Materials Science and Engineering, University of Maryland, College Park, Maryland 20742, USA}

\author{J. R. D. Copley}
\affiliation{NIST Center for Neutron Research, NIST, Gaithersburg, Maryland 20899-8102, USA}

\author{M. Ramazanoglu}
\affiliation{Department of Physics and
Astronomy, McMaster University, Hamilton, Ontario, L8S 4M1, Canada}

\author{H. A. Dabkowska}
\affiliation{Brockhouse Institute for Materials Research, Hamilton, Ontario, L8S 4M1, Canada}

\author{B. D. Gaulin}
\affiliation{Department of Physics and
Astronomy, McMaster University, Hamilton, Ontario, L8S 4M1, Canada}
\affiliation{Brockhouse Institute for Materials Research, Hamilton, Ontario, L8S 4M1, Canada}
\affiliation{Canadian Institute for Advanced Research, 180 Dundas St.\ W.,
Toronto, Ontario, M5G 1Z8, Canada}


\begin{abstract}

Co$_3$V$_2$O$_8$ is an orthorhombic magnet in which S=3/2 magnetic moments reside on two crystallographically inequivalent Co$^{2+}$ sites, which decorate a stacked, buckled version of the two dimensional kagome lattice, the stacked kagome staircase. The magnetic interactions between the Co$^{2+}$ moments in this structure lead to a complex magnetic phase diagram at low temperature, wherein it exhibits a series of five transitions below 11 K that ultimately culminate in a simple ferromagnetic ground state below T$\sim$6.2 K. Here we report magnetization measurements on single and polycrystalline samples of (Co$_{1-x}$Mg$_{x}$)$_{3}$V$_{2}$O$_{8}$ for $x$$<$0.23, as well as elastic and inelastic neutron scattering measurements on single crystals of magnetically dilute (Co$_{1-x}$Mg$_{x}$)$_{3}$V$_{2}$O$_{8}$ for $x$=0.029 and $x$=0.194, in which non-magnetic Mg$^{2+}$ ions substitute for magnetic Co$^{2+}$. We find that a dilution of 2.9$\%$ leads to a suppression of the ferromagnetic transition temperature by $\sim$15$\%$ while a dilution level of 19.4$\%$ is sufficient to destroy ferromagnetic long-range order in this material down to a temperature of at least 1.5 K. The magnetic excitation spectrum is characterized by two spin-wave branches in the ordered phase for (Co$_{1-x}$Mg$_{x}$)$_{3}$V$_{2}$O$_{8}$ ($x$=0.029), similar to that of the pure $x$=0 material, and by broad diffuse scattering at temperatures below 10 K in (Co$_{1-x}$Mg$_{x}$)$_{3}$V$_{2}$O$_{8}$ ($x$=0.194). Such a strong dependence of the transition temperatures to long range order in the presence of quenched non-magnetic impurities is consistent with two-dimensional physics driving the transitions.  We further provide a simple percolation model that semi-quantitatively explains the inability of this system to establish long-range magnetic order at the unusually-low dilution levels which we observe in our experiments. 

\end{abstract}
\maketitle

\section{Introduction}
Geometrically frustrated materials that are based on magnetic moments which decorate a lattice of triangular networks have been of great recent interest due to their intriguing low-temperature magnetic properties and a rich variety of exotic ground states such as spin glasses, spin ices, and spin liquids which they exhibit\cite{Ramirez_annurev,frustratedspinsystems}.

Co$_3$V$_2$O$_8$ belongs to the M$_3$V$_2$O$_8$ family (M=Co, Ni, Cu, Mn)\cite{Rogado02,RogadoCu3V2O8,Morosan} which incorporates into its structure an anisotropic variation of the ideal two dimensional (2D) kagome lattice. The kagome lattice is itself a network of corner-sharing triangles. Within this structure, the magnetic moments are carried by 3d transition metal M$^{2+}$ ions that decorate kagome layers. These layers buckle in and out of the plane, forming the kagome staircase structure. Co$_3$V$_2$O$_8$ (CVO) crystallizes in the orthorhombic space group Cmca\cite{Sauerbrei}. Edge-sharing Co$^{2+}$O$_6$ octahedra are situated slightly below and above the a-c plane, stacked along the crystallographic b-axis and are separated by nonmagnetic V$^{5+}$O$_4$ tetrahedra. A representation of the structure including only the Co$^{2+}$ ions is shown in Fig. 1a. The magnetic interactions between the Co$^{2+}$ S=3/2 moments are mediated by Co-O-Co superexchange, the resulting magnetic exchange pathways are indicated in Fig. 1a. as grey bonds. Within the buckled kagome layers that form the a-c plane (Fig. 1b.) the S=3/2 magnetic moments reside on two crystallographically inequivalent Co$^{2+}$ sites: crystallographic (8e) spine (s) sites (blue) run in chains along the a-axis, and these chains are linked by (4a) cross-tie (c) sites (red).

\begin{figure}[h]
\includegraphics{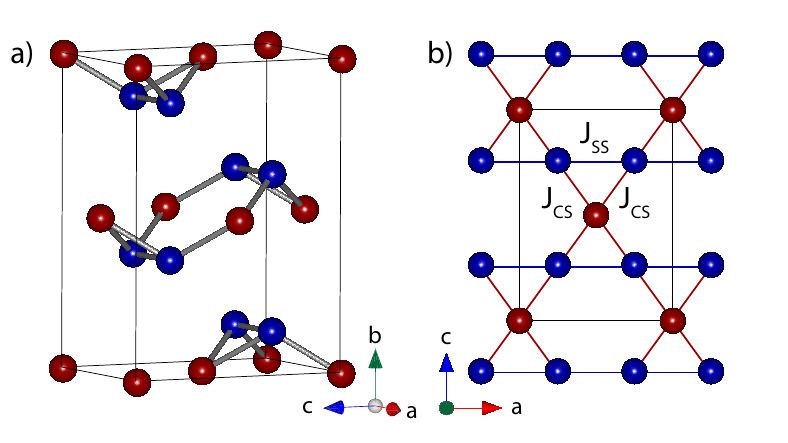}
\caption{The crystal structure of Co$_3$V$_2$O$_8$. a) View of the kagome staircase considering only the Co$^{2+}$ ions (red and blue) which are stacked along the \textit{b}-axis. b) View of the kagome plane projected on the \textit{a}-\textit{c} plane with the crystallographically inequivalent cross-tie (red) and spine sites (blue). The magnetic exchange interactions discussed in the text are indicated.}
\end{figure}

The low-temperature magnetic phase diagrams of all known members of the kagome staircase family are complex and show considerable diversity in their rich behavior, despite their isostructural nature\cite{Rogado02,RogadoCu3V2O8,Morosan}. For example, Ni$_3$V$_2$O$_8$ displays 4 different magnetically ordered states below 10 K and has attracted much interest due to the discovery of multiferroic behavior in one of its incommensurate phases at low temperature \cite{Kenzelmann06,LawesNi3V2O8_04,Rogado02}. The low-temperature phase diagram of Co$_3$V$_2$O$_8$ (CVO) has been studied by neutron diffraction in both zero\cite{Wilson07,Chen06,Mehmet09,Qureshi_form_factor09} and in finite applied magnetic fields\cite{Petrenko10,helton2012}. It has also been studied by optical spectroscopy\cite{Rai,Vergara_magnetoelastic_10}, by heat capacity and magnetization\cite{Rogado02,Szymczak06} as well as by $\mu$SR\cite{muons07} and by NMR techniques\cite{NMRCVO_NVO,VNMRCVO}. This extensive set of measurements has revealed a series of 5 different magnetic phase transitions below 11.2 K in zero applied magnetic field. These ultimately culminate in a simple ferromagnetic ground state below T$\sim$6.2 K. All five of the magnetic states display a preferred direction of the spins parallel to the \textit{a}-axis, the easy axis of this system. From previous inelastic neutron scattering measurements, the exchange parameters of the microscopic Hamiltonian in the ground state of CVO have been extracted to some approximation, using linear spin wave theory\cite{Mehmet09}. These calculations approximate the CVO system as two-dimensional and reveal that the exchange between magnetic moments on the cross-tie and spine sites is ferromagnetic and $J_{\mathrm{cs}}$$\sim$1.25meV, while the exchange between adjacent spine sites ($J_{\mathrm{ss}}$ in Fig. 1b) is negligible. Significant uniaxial anisotropy terms in the Hamiltonian have also been found ($\sim$1-2 meV), consistent with the easy a-direction.

Previous work has examined stacked kagome staircase magnets in the presence of disorder, specifically (Co$_{1-x}$Ni$_{x}$)$_{3}$V$_{2}$O$_{8}$, where random mixtures of Ni$^{2+}$ and Co$^{2+}$ reside on the M$^{2+}$ site\cite{Qureshi_Ni_Co_phasediagram,Qureshi_Ni_Co_magnetic_structure}. The phase behavior of such systems is expected to be complex, as both Ni$^{2+}$ and Co$^{2+}$ are magnetic, and a minimum of three exchange interactions would be necessary to describe the system, even if only a single M$^{2+}$ site was crystallographically relevant.

Here we report on (Co$_{1-x}$Mg$_{x}$)$_{3}$V$_{2}$O$_{8}$ which can be thought of in terms of Co$_3$V$_2$O$_8$ in the presence of quenched magnetic vacancies, as Mg$^{2+}$ does not carry a magnetic moment. As the ultimate ground state in Co$_3$V$_2$O$_8$ is a simple uniaxial ferromagnet, one would expect this to represent an excellent model for a quasi-two dimensional Ising system in the presence of quenched disorder. As such, one would expect to be able to understand its phase diagram and ground state properties in some detail.

Dilution studies of other Ising magnets have been carried out over a long period of time\cite{Birgeneau1}. In most cases, these can be understood in the context of percolation theory. However, there are interesting examples where this is not the case. One such example is the case of the stacked triangular lattice Ising-like antiferromagnet in the presence of quenched non-magnetic impurities, CsCo$_{1-x}$Mg$_{x}$Br$_3$, wherein the combination of antiferromagnetic exchange and triangular coordination leads the quenched magnetic vacancies to couple to the relevant order parameter as a random field\cite{CsCoMgBr_critical}.

In this paper, we present magnetization measurements on both polycrystalline and single crystalline samples of (Co$_{1-x}$Mg$_{x}$)$_{3}$V$_{2}$O$_{8}$ as well as elastic and inelastic neutron scattering measurements on single crystal (Co$_{1-x}$Mg$_{x}$)$_{3}$V$_{2}$O$_{8}$ with $x=0.029$ and $x=0.194$. We study the phase diagram as a function of Mg$^{2+}$ doping, $x$, and show how a simple two-dimensional percolation model can provide an explanation for the drastic suppression of the magnetic ordering at low temperatures in this material. In addition, we study the spin correlations and dynamics at low temperature as a function of doping, and show the existence of a Griffiths-like phase\cite{Griffiths} in the presence of quenched disorder, when T$_{C}(x)<$T$<T_{C}(x=0)$.

\section{Experimental Details}
Two large high-quality single crystals of (Co$_{1-x}$Mg$_{x}$)$_{3}$V$_{2}$O$_{8}$ with concentrations of $x$=0.029(3) and 0.194(4) were grown at McMaster University using an optical floating-zone image furnace\cite{Hanna}. Details of these growths and a subsequent X-ray structural refinement for these crystals are reported in \cite{CVOgrowth}. This work allows us to both quantify $x$, and to determine that the site dilution of the magnetic Co$^{2+}$ ions by non-magnetic Mg$^{2+}$ is almost random. The single crystals resulting from these growths had a mass of $\sim$ 8g and had approximately cylindrical shapes, with dimensions of 50 mm in length and 7 mm in diameter. Magnetization measurements were performed using a conventional SQUID magnetometer at McMaster University on several polycrystalline samples with a range of Mg-concentrations ($x=0-0.23$) as well as on three single crystals ($x$=0, $x$=0.029 and $x$=0.194). These single crystal samples were cut from the main crystal growths to a rectangular shape ($\sim$2.5 mm x 2 mm x 2 mm), and were aligned such that magnetization and susceptibility measurements could be performed with the magnetic field aligned along particular directions. All three single crystal samples used in these magnetization measurements had a mass of $\sim$10 mg. Preliminary neutron scattering data was acquired using the C5 triple-axis spectrometer at the Canadian Neutron Beam Centre (CNBC) at Chalk River Laboratories. Further neutron scattering measurements, using both triple-axis and time-of-flight techniques, were performed at the NIST Center for Neutron Research (NCNR). The experiments were performed on two $\sim$25 mm long cylindrical (Co$_{1-x}$Mg$_{x}$)$_{3}$V$_{2}$O$_{8}$ samples with $x$=0.029 and $x$=0.194, which were aligned such that the [H,0,L] plane in reciprocal space was coincident with the horizontal scattering plane.

Detailed elastic neutron scattering measurements were carried out using the cold triple-axis spectrometer SPINS at the NCNR, NIST. These measurements employed a vertically focussing PG-002 monochromator and flat PG-002 analyzer crystal with fixed final energy of $E_f$=5 meV. The collimation used was [open, 80', 80', 80'] with an in-pile, cooled Be filter placed in the neutron beam incident to the sample in order to eliminate higher-order wavelength contamination. Inelastic neutron scattering measurements were carried out on the time-of-flight Disk Chopper Spectrometer (DCS) at the NCNR, NIST employing a fixed incident wavelength of $\lambda_i$ = 2.5 $\AA$, which allowed for the measurement of magnetic excitations up to energy transfers of $\Delta$E $\sim$ 10 meV. In this configuration of DCS, a resolution of $\sim 0.9$ meV was obtained at the elastic position. The samples were placed in a conventional ILL Orange cryostat with a base temperature of 1.5 K for both the SPINS and DCS experiments.

\section{Magnetization Measurements}
We performed magnetization measurements using a SQUID magnetometer on a series of (Co$_{1-x}$Mg$_{x}$)$_{3}$V$_{2}$O$_{8}$ samples with different doping concentrations in the range ($x=0-0.23$). Measurements were carried out on polycrystalline samples as well as on single crystal samples ($x=0$, $x=0.029$ and $x=0.194$).
\begin{figure}[h]
\includegraphics{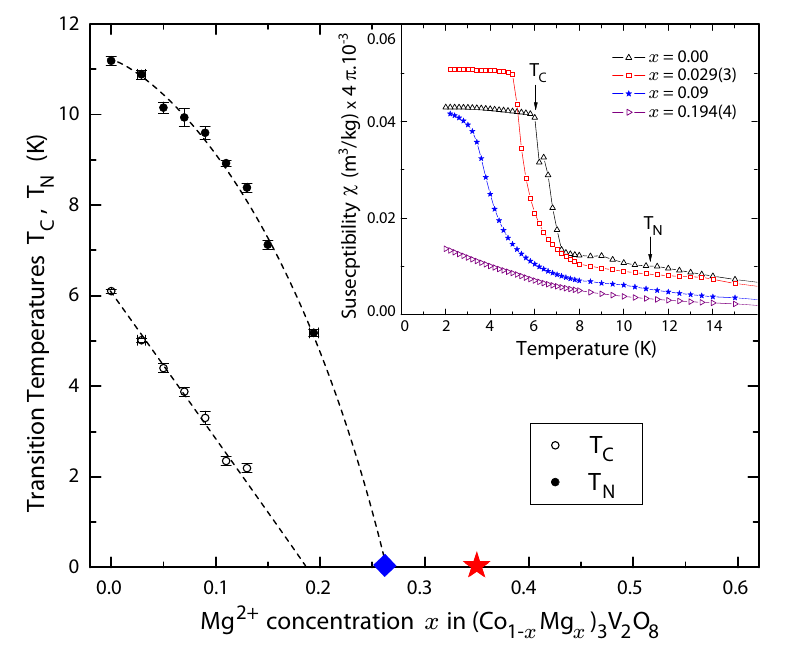}
\caption{(Color online) The phase diagram of (Co$_{1-x}$Mg$_{x}$)$_{3}$V$_{2}$O$_{8}$ as a function of $x$ obtained from magnetization measurements described in the text. The error bars result from the determination of the transition temperatures using data in Figure 3. The phase boundaries (dashed lines) are guides to the eye and are extrapolated to T = 0 K. The theoretically expected site percolation limit for the perfect 2D kagome lattice ($x_c$ $\sim$ 0.35)\cite{Sahimi94} is shown as the red star. The critical doping concentration $x_c$ above which any magnetic ordering gets suppressed at T=0 corresponds to the determined percolation limit discussed later in the text and is shown as blue diamond. The inset shows representative scans of the field-cooled mass susceptibility $\chi(\mathrm{T})$ for an applied field of $\mu_0$H=0.005 T, and the phase transition temperatures T$_\mathrm{C}$ and T$_\mathrm{N}$ for data points in the phase diagram.}
\end{figure}

The temperature-dependent mass susceptibility $\chi(\mathrm{T})$ was measured in a field-cooled (FC) mode in an applied field of $\mu_0$H=0.005 T and this is shown in the inset of Fig. 2 for selected values of $x$. For the measurements on single crystals (open symbols), the crystals were aligned with their a-axis parallel to the applied magnetic field. Measurements on pure Co$_3$V$_2$O$_8$ show two well-defined magnetic transitions at T$_\mathrm{N}\sim11.2$ K and at T$_\mathrm{C}\sim$6.2 K as indicated by arrows in the inset. We associate the higher temperature transition with a transition from the paramagnetic to an incommensurate antiferromagnetic phase, and the lower temperature transition at $\sim$6.2 K with the transition to the ferromagnetic ground state, as has been reported previously \cite{Chen06}. These two transition temperatures were extracted from $\chi(\mathrm{T})\cdot \mathrm{T}$ curves, shown for a subset of the data in Fig. 3, resulting in the phase diagram shown in Fig. 2.

\begin{figure}[h]
\includegraphics{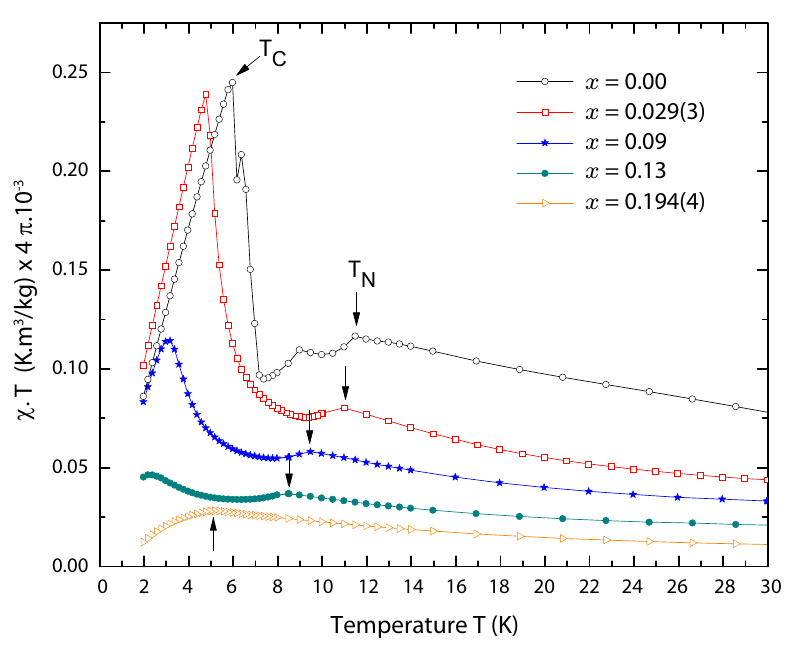}
\caption{(Color online) Representative scans of the susceptibility $\chi(\mathrm{T})\cdot T$ reveal anomalies at the phase transition temperatures T$_\mathrm{C}$ and T$_\mathrm{N}$ (indicated by arrows). These transition temperatures are used to construct the phase diagram shown in Fig. 2. Closed symbols refer to data obtained from polycrystalline samples, while open symbols represent measurements on single crystals. Note that the maximum in the $\chi(\mathrm{T})\cdot T$ curve in the x=0.194 doped sample is indicated as phase transition with a T$_\mathrm{N}\sim$ 5 K in Fig. 2, however, elastic neutron scattering data shows no clear evidence for long range magnetic correlations at any temperature above 1.5 K in this sample.}
\end{figure}

These measurements reveal a suppression of both transition temperatures with increasing $x$, and the complete suppression of any magnetic ordering at T=0 can be extrapolated to a critical doping concentration of $x_c\sim$0.26. This result is somewhat unexpected since this doping level is lower than the theoretical percolation threshold for the destruction of long-range order for the ideal 2D kagome lattice, for which  $x_c\sim$0.35 (indicated as red star in the phase diagram) \cite{Sykes,Sahimi94}. It is also much lower than the percolation threshold for any three dimensionally ordered system where $x_c\sim$0.68-0.80 \cite{Essam,Stauffer1979,Gaunt}. Note that the percolation threshold $p_c$ reported in the literature is related to $x_c$ discussed here by $x_c=1-p_c$. While $p_c$ describes the critical concentration of {\it magnetic} ions below which magnetic long range order ceases to exist, $x_c$ is used for the complementary description of the critical concentration of {\it non-magnetic} ions above which long range order is impossible. We choose to speak of the percolation threshold $x_c$ in this paper instead of $p_c$ as it is directly related to the concentration of non-magnetic Mg$^{2+}$ ions in our doped CVO.

From the phase diagram in Fig. 2, it can be seen that the ferromagnetic transition temperature T$_\mathrm{C}$ is very sensitive to magnetic dilution and changes almost linearly with $x$. It is lowered by $\sim$20$\%$ from T$_\mathrm{C}$$\sim$6.2 K in the pure material to T$_\mathrm{C}$$\sim$5.2 K for a doping level of $x$=0.029, and for $x$=0.194 it is suppressed below the lowest accessible temperature for our SQUID measurements (T$\sim$2 K). In contrast, the higher temperature Neel transition changes more gradually with dilution, exhibiting downwards curvature as a function of $x$. The reason for this difference is not clear, although we do note that while the transition at T$_N$ appears to be continuous, that at T$_C$ is clearly discontinuous even in the presence of quenched disorder, as we will discuss below.

\section{Elastic Neutron Scattering Results}
Elastic neutron scattering measurements were performed on two doped single crystal samples ($x$=0.029 and $x$=0.194) using the cold triple-axis spectrometer SPINS at the NCNR, NIST. The temperature dependence of the elastic magnetic scattering was followed around the (0,0,2) Bragg peak, which exhibits a weak nuclear component, and a strong ferromagnetic component to the Bragg peak below T$_C$$\sim$6.2 K in the pure, $x$=0, material. The strength of the magnetic Bragg peak follows as a consequence of the form of the magnetic neutron scattering cross section which is sensitive to the component of moment perpendicular to ${\bf Q}$. Its strength indicates that the ordered moment points along {\bf a}. A color contour plot constructed from elastic longitudinal (0,0,$L$) scans through (0,0,2) as a function of temperature for both samples is presented in Fig. 4 a) and c). Representative scans are shown in Fig. 4 b) and d). Note that the intensity appropriate to the color contour plots in Fig. 4 a) and c) is on a log scale.

\begin{figure}[h]
\includegraphics{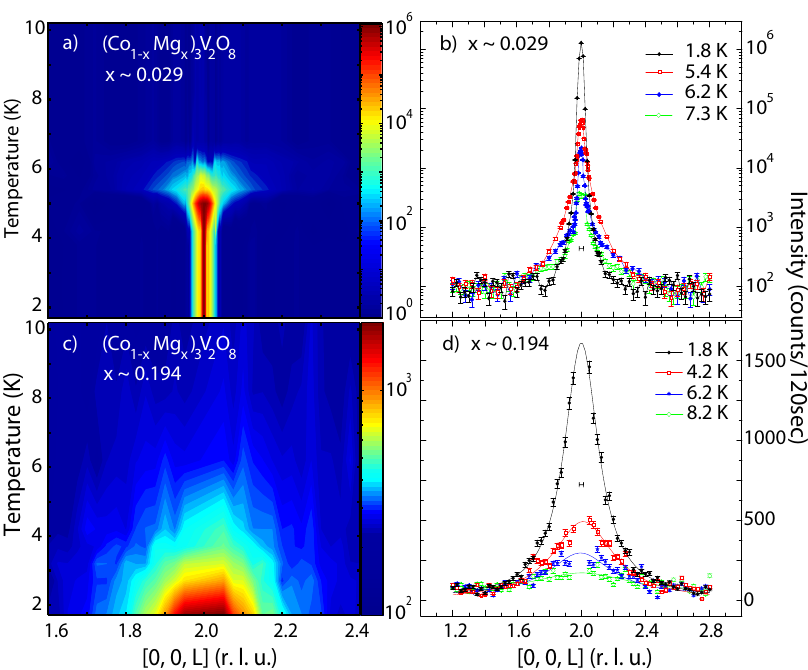}
\caption{(Color online) Contour plots showing the temperature evolution
of the elastic magnetic scattering S({\bf Q}, E=0) around the {\bf Q}=(0,0,2) Bragg peak position for a) $x$ = 0.029 and c) $x$ = 0.194. A background data set within the paramagnetic state at T = 20 K has been subtracted from the individual (0,0,L) scans so as to eliminate the weak nuclear component to the (0,0,2) Bragg scattering. Representative scans making up the contour plots are shown in panels b) and d). The instrumental resolution of SPINS is indicated in panels b) and d) as the black horizontal bar. The error bars represent one standard deviation in this and the following figures.}
\end{figure}


In the data sets for both $x$=0.029 and $x$=0.194, a high temperature data set at T = 20 K (in the paramagnetic phase) has been subtracted, so as to eliminate the temperature-independent, nuclear scattering from the lower temperature data sets, and hence to isolate the magnetic scattering. It can easily be seen that the scattering profiles for the two samples are in striking contrast to each other. As the temperature is lowered below T$\sim$T$_{C}(x=0)\sim$ 6.2 K in the $x$=0.029 Mg-doped sample (Fig. 4 a) and b)), diffuse scattering intensity characteristic of the development of short range correlations builds up, increasing in strength down to a temperature of T$_{C}(x=0.029)\sim$5.2 K. Below $\sim$5.2 K it gives way to a dramatic increase in elastic scattering intensity which is now sharper in {\bf Q} and characteristic of long range order.  It is therefore associated with a phase transition to the ferromagnetic ground state. This can be seen more clearly in Fig. 5, which displays the temperature dependence of the net scattered intensity upon warming and cooling measured at the Bragg peak position (0,0,2) associated with the order parameter, and at a {\bf Q} position slightly away from the Bragg peak, at (0,0,1.8), allowing for the parametrization of the diffuse scattering characteristic of the short range order. Based on the sharpness of the transition and the observed hysteresis in the order parameter, this transition is clearly of first-order nature.

\begin{figure}[h]
\includegraphics{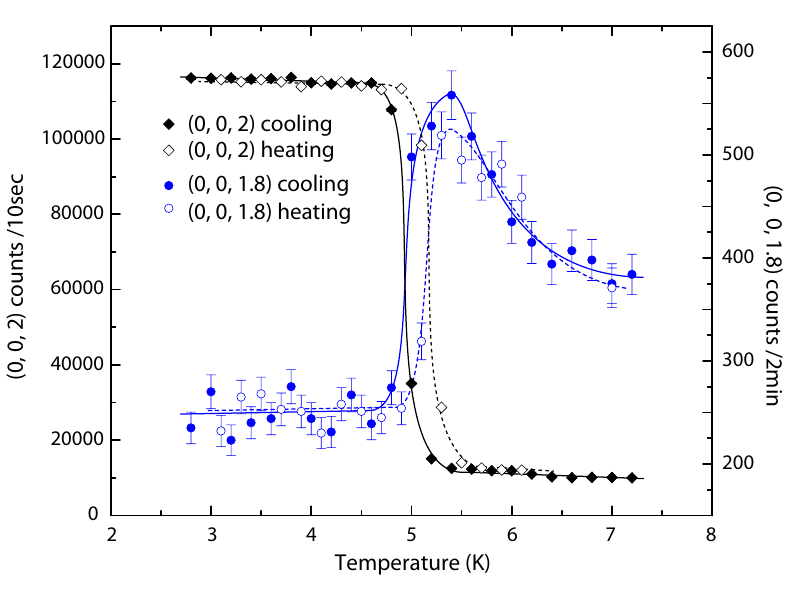}%
\caption{The order parameter and diffuse scattering as a function of temperature measured near the (0,0,2) magnetic Bragg peak in the $x$=0.029 sample. Note the different intensity scales and counting times.}%
\end{figure}

The diffuse critical scattering in the wings of the elastic Bragg peak is observed to be drastically enhanced in a Griffiths-phase-like fashion\cite{Griffiths, Bray} in the temperature region between T$_{C}(x=0.029)\sim$5.2 K and T $<$T$_{C}(x=0)\sim$6.2 K. Such a Griffiths phase shows enhanced spin correlations due to relatively rare, large correlated spin droplets\cite{Fisher} that arise as a consequence of the random quenched disorder\cite{Griffiths}. The effect of thermal fluctuations is such that the system would be long range ordered in the absence of quenched disorder, and large percolating networks of spins display enhanced correlations somewhat akin to long range order.

The (0,0,$L$) scans of the magnetic scattering from this $x$=0.029 sample are best described by a two-component lineshape consisting of a Lorentzian lineshape characterizing the short-range correlations and a resolution-limited Gaussian lineshape which describes the onset of long-range magnetic order.

The results of this lineshape analysis are given in Fig. 6, wherein the integrated intensities of the Gaussian and Lorentzian components of the magnetic scattering at (0,0,2) for the $x$=0.029 sample are shown in panel a). A representative elastic scan with corresponding fits at T = 5.8 K, in the Griffiths-like phase, between T$_C(x=0)$ and T$_C(x=0.029)$ is displayed in the inset of panel a).
\begin{figure}[h]
\includegraphics{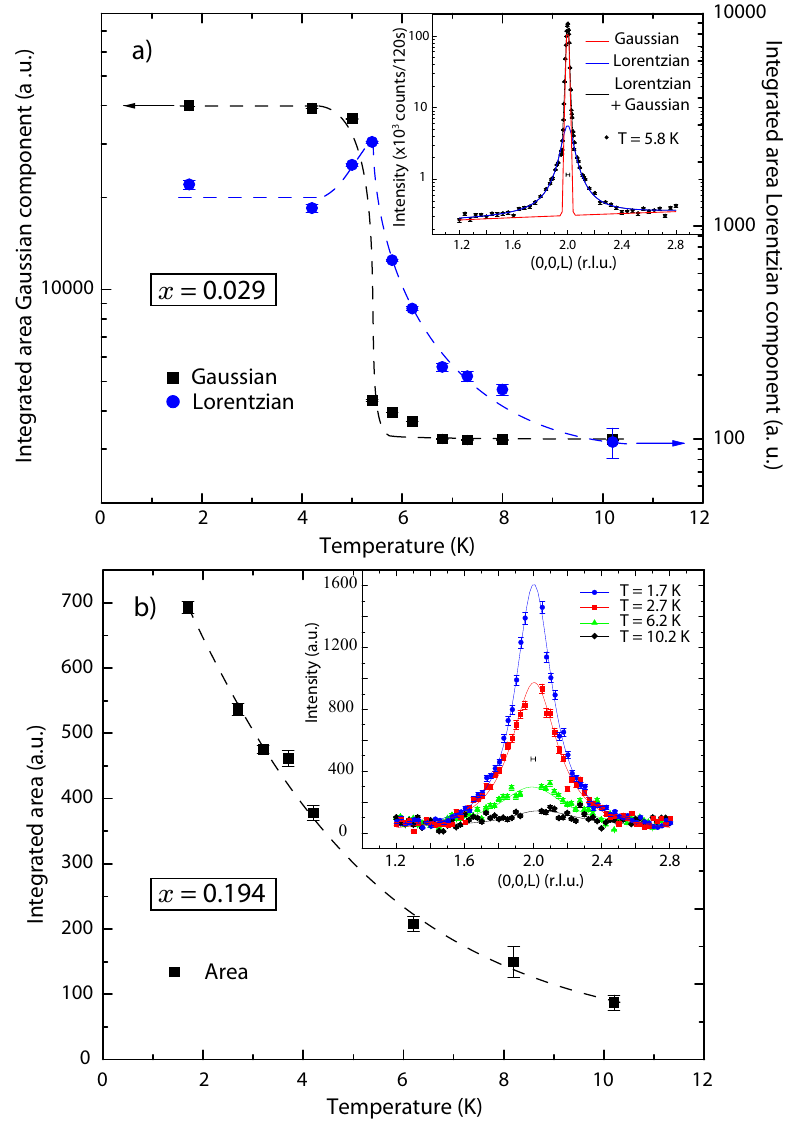}
\caption{(Color online) Elastic scattering in the a) $x$=0.029 and b) $x$=0.194 doped (Co$_{1-x}$Mg$_{x}$)$_{3}$V$_{2}$O$_{8}$ samples. a) The integrated intensities of the Gaussian (black) and Lorentzian (blue) components to the lineshape used to describe the magnetic scattering in the $x$=0.029 sample discussed in the text are shown. A representative scan with corresponding fits in the Griffiths-like phase at T = 5.8 K is shown in the inset. b) The integrated intensity of the Lorentzian lineshape used to describe the magnetic scattering in the $x$=0.194 doped sample is shown, while the inset displays representative scans in L along the (0, 0, 2) direction for different temperatures with fits to the Lorentzian lineshape as described in the text. For both panels a) and b) a T = 20 K high temperature non-magnetic background has been subtracted to isolate the magnetic scattering. Note that the peaks for the $x$=0.194 doped sample in the inset of panel b) are much wider than the instrumental resolution at SPINS, which is shown as black horizontal bar.}
\end{figure}
For the higher doping level of the $x$=0.194 sample, the scattering profile is in sharp contrast to that at $x$=0.029, as shown in the color contour maps in Fig. 4 and with the data we now discuss in Fig. 6. In agreement with the magnetization data, no magnetic long range order is evident down to the lowest temperature measured, $\sim$1.5 K. Instead, the scattering profile is entirely dominated by a broad diffuse scattering component around the (0,0,2) Bragg position, indicative of very short-ranged correlations. After subtraction of the high-temperature nuclear Bragg intensity, it was found that a Lorentzian lineshape alone was appropriate to describe the magnetic scattering profile at all temperatures. Figure 6 b) shows the integrated intensity of the magnetic diffuse scattering around (0,0,2) and the inset shows several representative scans and the corresponding Lorentzian fits to this scattering for $x$=0.194. One can clearly see that the diffuse scattering is very weak at 10 K and that it develops noticeably below T$_{C}(x=0)\sim$6.2 K, again consistent with a Griffiths-like phase between T$_C(x=0)$ and zero temperature, increasing monotonically on cooling down to 1.5 K. 

\begin{figure}[h]
\includegraphics{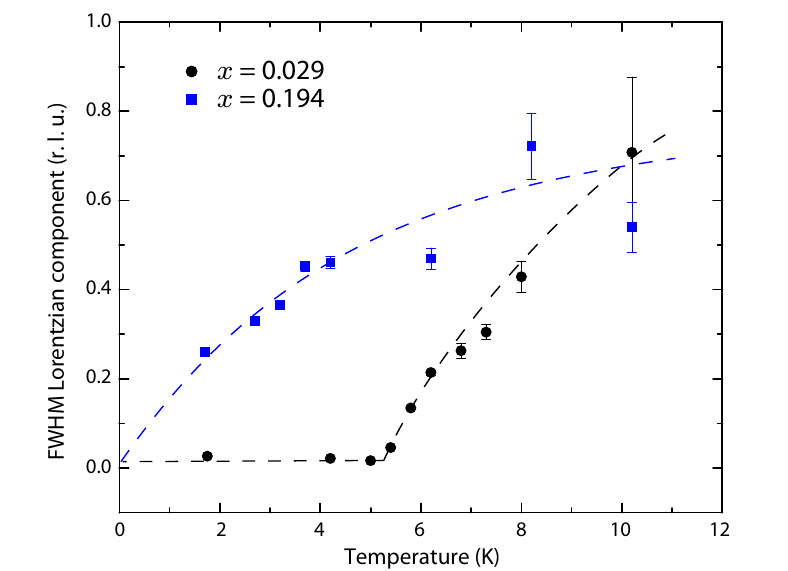}
\caption{(Color online) FWHM of the Lorentzian components shown for both doped single crystal samples. The dashed lines are guides to the eye. While for $x$=0.029, the FWHM of the scattering near (0,0,2) drops to 0 at 5.2 K as the sample develops long-range order, for the higher doping of $x$=0.194, the scattering maintains a finite {\bf Q}-width to the lowest temperatures measured. The trend for the evolution of the FWHM for $x$=0.194 however, extrapolates approximately to 0 at T=0. }
\end{figure}

The full widths at half maximum (FWHM) of the Lorentzian components characterizing the short-range correlations in both the (Co$_{1-x}$Mg$_{x}$)$_{3}$V$_{2}$O$_{8}$ samples with $x$=0.029 and $x$=0.194 are compared in Fig. 7. It is clear that the FWHM of the x=0.029 sample drops to 0 at T$_C(x=0.029)\sim$ 5.2 K where the sample develops long range order (as evidenced by a resolution-limited Gaussian lineshape), whereas magnetic scattering from the $x$=0.194 sample maintains a finite width to the lowest temperatures measured, T = 1.5 K.  At this temperature, the (0,0,2) peak displays a FWHM of $\sim$0.25, corresponding to a real-space spin-correlated-droplet within the kagome plane of about 24 $\AA$ in diameter, approximately the size of two to three unit cells. While no phase transition to long-range order is observed for $x$=0.194, the trend for the evolution of the FWHM of this magnetic scattering approximately extrapolates to 0 at T = 0, indicating that $x$=0.194 is close to a quantum critical point. We also note that the FWHM for both samples approach each other at the highest temperature measured $\sim$10 K, as we anticipate should occur. At these high temperatures, thermal fluctuations are expected to dominate the effects of the quenched disorder.

\begin{figure}[h]
\includegraphics[width=8.8cm]{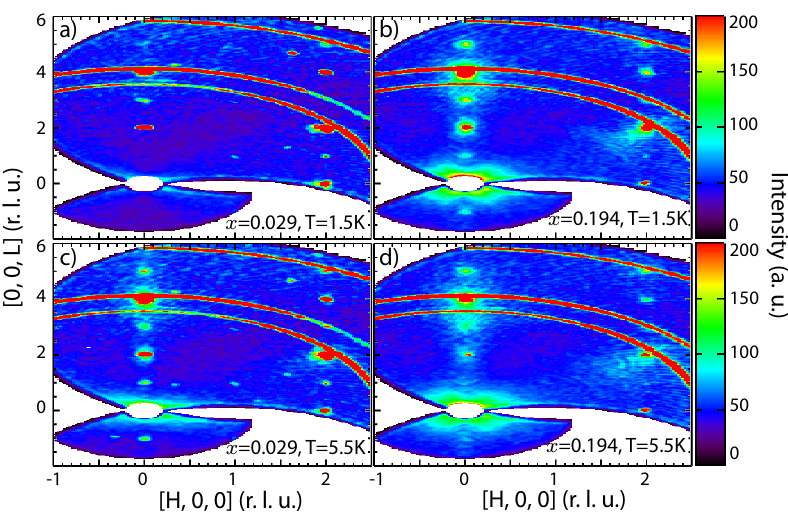}
\caption{(Color online) Maps of the elastic scattering intensity S({\bf Q}, E = 0) for (Co$_{1-x}$Mg$_{x}$)$_{3}$V$_{2}$O$_{8}$ with x=0.029 and x=0.194 and for two different temperatures, T = 1.5 K (top row) and T = 5.5 K (bottom row).}
\end{figure}

We further investigated the nature of the elastic scattering using TOF neutron data taken on DCS at NCNR, NIST. The obtained maps of elastic scattering S({\bf Q}, E=0) are shown in Fig. 8 a), c) for $x$=0.029 and in Fig. 8 b) and d) for $x$=0.194, each for two different temperatures, T = 1.5 K and T =5.5 K. 
\begin{figure}
\includegraphics{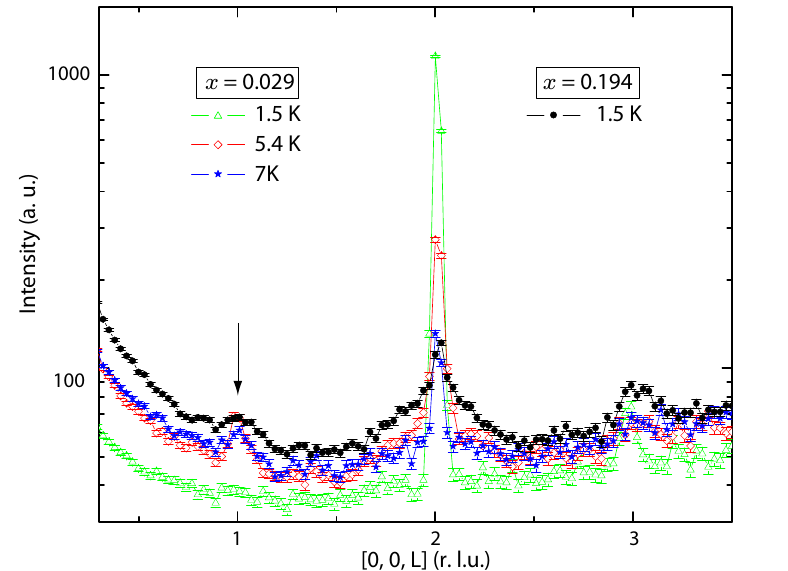}
\caption{(Color online) Cuts through the color contour maps maps shown in Fig. 8 along 0,0,L. The observed (0,0,1) peak at T = 1.5 K in the $x$=0.194 sample is a clear signature of incommensurate fluctuations of the form (0,$\delta$,L) that are observed in the $x$=0.029 sample at T = 5.5 K and T = 7 K. This is consistent with the phase diagram presented in Fig. 2.}
\end{figure}
One notes that the net elastic scattering at T = 1.5 K, following subtraction of a T = 20 K data set, is different for the $x$=0.194 sample than for either the pure $x$=0 (not shown here), or the lightly doped $x$=0.029 sample. Figure 9 shows that Bragg-like scattering is observed for the $x$=0.194 sample around all of the (0,0,L) Bragg positions for integer L, likely originating from short range incommensurate elastic magnetic scattering of the form (0, $\delta$, L). This scattering, which does not lie in our horizontal scattering plane, is picked up in our DCS experiment due to the finite acceptance of the scattered neutrons out of the horizontal plane, and is consistent with the placement of the $x$=0.194 single crystal sample on the phase diagram for (Co$_{1-x}$Mg$_{x}$)$_{3}$V$_{2}$O$_{8}$ as shown in Fig. 2.

\section{Inelastic Neutron Scattering Results}

We investigated the evolution of the dynamic spin correlations in (Co$_{1-x}$Mg$_{x}$)$_{3}$V$_{2}$O$_{8}$, the excitations out of the ground state, using inelastic neutron scattering measurements carried out on DCS at NCNR, NIST. Figure 10 shows S({\bf Q}, E) observed in the three $x$=0, $x$=0.029, and $x$=0.194 single crystal samples along two perpendicular directions in reciprocal space.

\begin{figure}[h]
\includegraphics{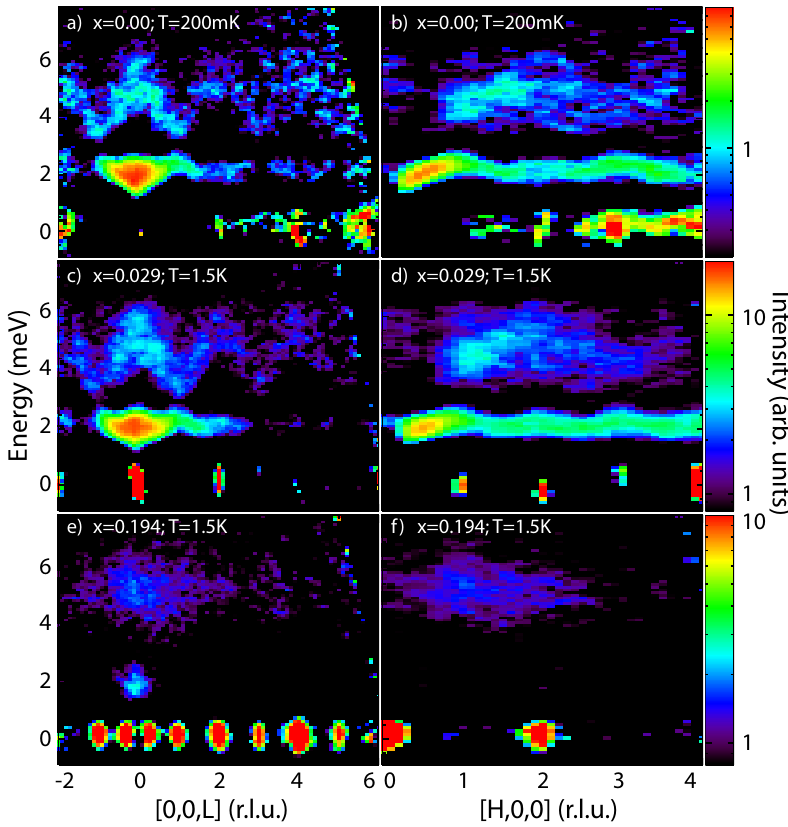}
\caption{(Color online) S({\bf Q},E) in $x$=0 (top), $x$=0.029 (middle) and $x$=0.194 (bottom row) samples along two perpendicular directions in reciprocal space. All data shown has had its respective high temperature T = 20 K background data set subtracted. Note that the data has been taken at the base temperatures appropriate to each experiment, which was 200 mK for $x$=0 and T = 1.5 K for both $x$=0.029 and $x$=0.194.}
\end{figure}

Panels a) and b) in the top row of Fig. 10 show S({\bf Q},E) for the pure, x=0, sample at T = 200 mK, panels c) and d) correspond to the $x$=0.029 sample at T = 1.5 K, and e) and f) correspond to $x$=0.194. The respective 20 K, high temperature data sets for each $x$, have been subtracted for all data sets shown. The spin wave spectrum in all three samples consists of two bands of spin excitations, a relatively narrow band near $\sim$ 2 meV and a broader band between $\sim$ 3 and 6.5 meV. These are consistent with earlier spin wave results on the pure $x$=0 sample\cite{Mehmet09}.

\begin{figure}[h]
\includegraphics{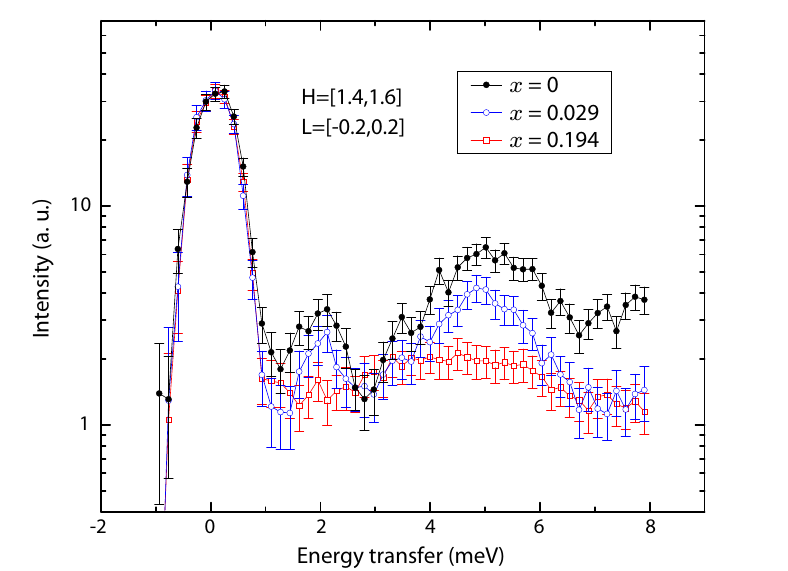}
\caption{(Color online) Cuts through scattering intensity around [1.5,0,0] and integrating over a narrow range in H=[1.4,1.6] and L=[-0.2,0.2] on a logarithmic intensity scale. To facilitate the comparison between these two sets of measurements, the scattered intensity has been normalized to the incoherent elastic scattering of the pure sample.}
\end{figure}

As expected from the elastic neutron scattering results discussed above, the inelastic spectrum for the pure and the $x$=0.029 samples look similar within their ordered states, although broadening of the spin wave excitations in energy is clear on doping, corresponding to shorter spin wave lifetimes in the presence of quenched impurites for the $x$=0.029 sample. The inelastic scattering from the most heavily doped $x$=0.194 sample shows poorly defined bands of spin excitations which are most easily observable in the vicinity of {\bf Q}=0. These are typical of the magnetic inelastic spectrum within the incommensurate phases or paramagnetic state at elevated temperatures.

The inelastic magnetic spectrum can be further compared between the three $x$=0, $x$=0.029, and $x$=0.194 samples, by taking cuts through the color contour maps of S(${\bf Q}$,E) shown in Fig. 10, and normalizing to the incoherent elastic scattering. This normalization should account for the somewhat different volumes of sample within the neutron beam. This is shown in Fig. 11, where we have shown an approximate constant-${\bf Q}$=(1.5, 0, 0) scan, which has been constructed by integrating in H between 1.4 and 1.6 and in L between $\pm$ 0.2. The spectral weight of the sharp spin excitations falls off dramatically with increased concentration of quenched non-magnetic disorder.

\begin{figure}[h]
\includegraphics[width=8.5cm]{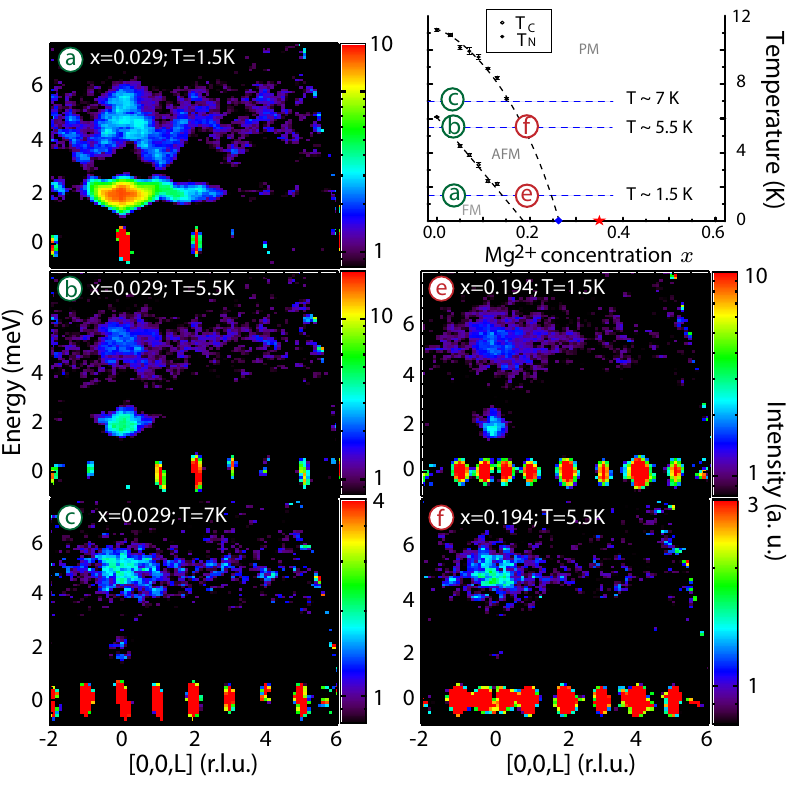}
\caption{(Color online) The temperature evolution of the inelastic scattering S({\bf Q},E) along [0,0,L], contrasting the $x$=0.029 and $x$=0.194 samples. Panels a), b) and c) correspond to the $x$=0.029 sample at the temperatures indicated in phase diagram in the top right panel. Panels e) and f) correspond to the $x$=0.194 sample with temperatures as shown in the phase diagram.}
\end{figure}

Further, we investigated the temperature evolution of the inelastic scattering in the two doped samples, $x$=0.029 and 0.194, as shown in Fig. 12 which shows S({\bf Q},E) along the [0,0,L] direction in reciprocal space. This is consistent with our elastic scattering studies, which placed the $x$=0.194 doped sample at 1.5 K in a region of the phase diagram that is characterized by incommensurate elastic magnetic order and rather diffuse spin wave excitations.  This corresponds to the same region of the phase diagram in which one finds the $x$=0.029 sample at elevated (T = 5.5 K and T = 7 K) temperatures. Note also the ressemblance of b) and c) with e), which are all placed in the incommensurate antiferromagnetic phase in the phase diagram of diluted CVO.

\section{Percolation Calculations and Discussion}

In light of the phase diagram obtained for (Co$_{1-x}$Mg$_{x}$)$_{3}$V$_{2}$O$_{8}$, shown in Fig. 2, and both the elastic and inelastic neutron scattering in the presence of quenched disorder, one can raise the question as to why dilution with quenched magnetic vacancies is so effective in destroying the magnetic order in this system. Figure 2 shows the ferromagnetic transition to be suppressed to zero temperature at $x\sim$0.19, while all vestiges of incommensurate magnetic order have been suppressed to zero temperature by $x\sim$0.26. This is far below the percolation threshold for any three dimensional cooperative system, where critical concentrations of $x_c\sim0.7$ ($p_c\sim0.3$) are typical\cite{Gaunt,Uemura}.
In and of itself, this shows that the weak three dimensional interactions along the third, stacking dimension, {\bf b}, do not determine this criticality. These interactions are certainly weak compared to the strong interactions within the kagome-staircase plane, but they may also be frustrated as a consequence of the triangular coordination of spins on cross-tie sites which neighbor the spine-site spins along the {\bf b}-axis. 


To further examine this question, we carried out a simple two dimensional, zero temperature percolation calculation relevant to (Co$_{1-x}$Mg$_{x}$)$_{3}$V$_{2}$O$_{8}$. Our starting point for a calculation of the percolation limit in this system was a two-dimensional projection of the magnetic moments within a single buckled kagome layer. For the calculations, a lattice of 100x100 unit cells was used, from which we randomly removed magnetic sites (either cross-tie or spine positions) with probability $x$. As Ramazanoglu et al.'s \cite{Mehmet09} linear spin wave theory on the pure CVO material showed the magnetic interactions between magnetic moments on the spine sites to be negligible, we then \textit{excluded} magnetic interaction pathways between spine sites for the remaining steps in the calculation. We calculated the number of sites contained in the largest cluster, that is the number of sites contained in the largest near-neighbor connected region of the lattice, and checked whether any continuously connected path through the lattice (chosen arbitrarily from the left to the right edge of the lattice) could be found. If several such connected paths were found, the length of the shortest path through the lattice between sites making up the largest lattice-spanning cluster was calculated. We ran this simulation up to 30 times for each dilution probability $x$ to get a good estimate of the percolation threshold, which we find to be at $x_c$$\sim$0.26. A calculation for the full connectivity of the site-diluted lattice was performed as well, by \textit{including} the interaction pathways between spine sites. This calculation recovers the well-known percolation threshold for the two-dimensional kagome lattice of $x_c$$\sim$0.35 ($p_c$=0.65). Though the finite size of the system smears out the actual percolation threshold $x_c$, the threshold for the original kagome lattice is well reproduced\cite{Sahimi94}. Increasing the lattice size does not change the result for $x_c$ significantly, and we therefore conclude that our estimate of $x_c\sim0.26$ for the case of (Co$_{1-x}$Mg$_{x}$)$_{3}$V$_{2}$O$_{8}$, where the magnetic interactions between spine sites is neglible, is accurate.

\begin{figure}[hhh]
\includegraphics[width=8cm]{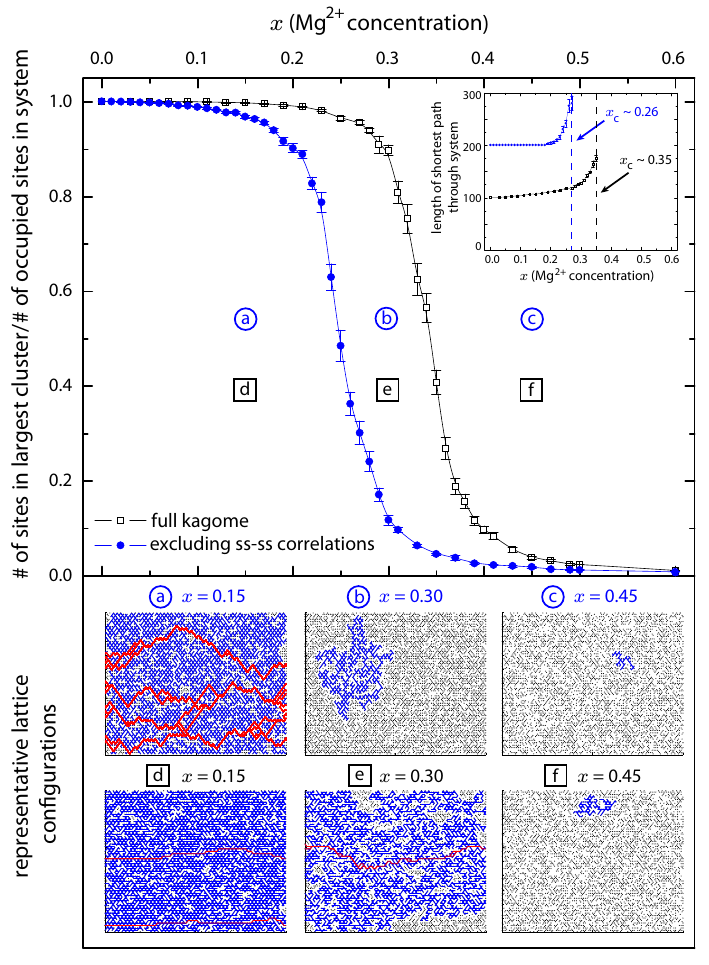}
\caption{(Color online) Percolation calculations showing the evolution of the largest cluster size in a lattice containing 100x100 unit cells as a function of site dilution $x$. The inset shows the length of the shortest path through the system connecting the edges of the lattice, if it exists. Above the percolation threshold, the lattice edges cannot be connected, and there is no shortest path. Note that the evolution of the largest cluster size is normalized to its value at zero site dilution ($x$=0). Error bars in the normalized largest cluster sizes are determined from the standard error. The determination of the shortest path through the system was based on only those data sets that showed connectivity of the lattice, and the error bars are given by the standard error of these data sets.}
\end{figure}

A summary of both simulations is shown in Fig. 13. In the top panel, the number of sites within the largest cluster, normalized to the total number of sites occupied by magnetic moments (which depend on $x$), is shown. The black squares denote the data for the case of full kagome connectivity, while data in blue circles represents the case in which spine-spine connectivity (ss-ss correlations) is excluded, corresponding to the physical picture relevant to (Co$_{1-x}$Mg$_{x}$)$_{3}$V$_{2}$O$_{8}$. In the inset, the length of the shortest path, if it exists, is shown for both cases. The point $x_c$ for which the lattice cannot be connected and thus no shortest path can be found, is indicated by a dashed line and is determined as $x_c\sim0.35$ for the full kagome connectivity and as $x_c\sim0.26$ for the case appropriate to (Co$_{1-x}$Mg$_{x}$)$_{3}$V$_{2}$O$_{8}$. It is interesting to note that the length of the shortest path increases relatively gradually for the full kagome connectivity, while excluding spine-spine connections leads to a flat dependence on the length of the shortest path through the system, until a dilution level of about $x\sim0.20$ is reached, beyond which the length of the shorest path rises rapidly towards $x_c\sim0.26$. This seems reasonable as the random removal of cross-tie and spine sites is equivalent in the fully connected kagome case, while the removal of cross-tie sites, which make up 1/3 of all sites is more effective than the removal of the majority spine sites, for the case where magnetic interactions between spine sites is negligible. This is because removal of a crosstie site ``disables'' a possible path to 4 spine sites, while removal on a spine site only ``disables'' the interaction path between 2 crosstie sites.

Representative lattice configurations at dilution levels of $x$=0.15, $x$=0.30 and $x$=0.45 are shown in the bottom panel of Fig. 13, for both cases. The largest cluster in the lattice is colored in blue and the shortest lattice spanning path(s) are shown in red. As one can clearly see, at a dilution level of $x=0.30$, there exists a largest, lattice spanning cluster in the case of full kagome connectivity (bottom middle panel), while for the case of excluding spine-spine connectivity as is relevant for (Co$_{1-x}$Mg$_{x}$)$_{3}$V$_{2}$O$_{8}$, only a finite-sized cluster exists, which does not span the full lattice and thus does not allow for long range magnetic order.


\section{Conclusions}
We have presented the low temperature phase diagram of the kagome staircase system CVO in the presence of quenched disorder, showing the transition temperatures T$_N$ and T$_C$ as function of non-magnetic Mg$^{2+}$ doping in (Co$_{1-x}$Mg$_x$)$_3$V$_2$O$_8$. We have found that the magnetic properties of this material at low temperatures are very susceptible to quenched disorder and that a doping level of $x\approx0.26$ is large enough to suppress any long range order to below 1.5 K.

Based on the geometric arrangement of the magnetic Co$^{2+}$ moments within the kagome plane, a simple two-dimensional percolation model has been used to understand the effect of magnetic dilution on the possible exchange paths. For that purpose, we considered a flat, two-dimensional kagome lattice with site dilution $x$, and calculated the percolation threshold. Based on previous inelastic neutron scattering on CVO, which found the spine-spine magnetic interactions to be negligible compared with the crosstie-spine interactions\cite{Mehmet09}, we calculated the percolation threshold of $x_c\sim0.26$ for the case of crosstie-spine connectivity (excluding spine-spine bonds). As a reference, we confirmed the percolation threshold for the kagome lattice as $x_c\sim0.35$ ($p_c\sim$0.65). The calculation of the percolation threshold excluding spine-spine connectivity is found to be in good agreement with the observed phase diagram that puts the percolation threshold at $\sim$26$\%$ percent doping level.

For two samples with doping levels of $x$=0.029 and $x$=0.194, respectively, we have performed elastic and inelastic neutron scattering measurements. The elastic neutron scattering measurements could be understood in terms of Griffiths-like phase fluctuations appearing between T$_C(x=0)$ and T$_C(x)$, wherein enhanced short range order is observed. For the case of the lightly doped $x$=0.029 sample, such short range correlations collapse into a long range ordered ferromagnetic state via a first order phase transition at T$_C$$\sim$5.2 K. The more heavily doped $x$=0.194 sample displays only ferromagnetic short range order to the lowest temperatures measured, and shows
the co-existence of incommensurate fluctuations at base temperature, consistent with the phase diagram for (Co$_{1-x}$Mg$_{x}$)$_{3}$V$_{2}$O$_{8}$ described above.

Our inelastic neutron scattering measurements on the lightly-doped $x$=0.029 sample show spin waves comparable to those observed in pure, $x=0$, CVO, albeit with finite energy widths, and hence spin wave lifetimes within the ground state. The equivalent magnetic excitation spectrum observed in the $x$=0.194 sample shows magnetic spectral weight in the same energy regime, but no well defined spin wave excitations at any wavevevctor.

\section*{Acknowledgments}
This work utilized facilities supported in part by the National Science Foundation under Agreement
No. DMR-0944772, and was supported by NSERC of Canada. The DAVE software package\cite{DAVE} was used for data reduction and analysis of DCS data. The authors would like to thank A. B. Dabkowski and C. Majerrison for assistance with sample preparation and P. Dube for assistance with the magnetization measurements.

%
\end{document}